\documentclass[journal]{IEEEtran}

\usepackage{grffile}
\usepackage{cite}
\usepackage[pdftex]{graphicx}
\usepackage{epstopdf}
\usepackage{amsmath}
\usepackage{amssymb}
\usepackage[english]{babel}
\usepackage{fixltx2e}
\usepackage{dsfont}
\usepackage{color}
\usepackage{algorithm}
\usepackage{algpseudocode}
\usepackage{textcomp}
\usepackage{balance}

\ifCLASSINFOpdf
\else
\fi

    \usepackage{ifpdf}

\begin{document}

\title{Robust Recovery of Missing Data in Electricity Distribution Systems}

\author{	
Cristian Genes,
I\~{n}aki~Esnaola, 
Samir~M.~Perlaza, 
Luis F. Ochoa,
and Daniel Coca.

\thanks{Cristian Genes and Daniel Coca are with the Department of Automatic Control and Systems Engineering, University of Sheffield, Sheffield S1 3JD, UK.}
\thanks{ I\~{n}aki Esnaola is with the Department of Automatic Control and Systems Engineering, University of Sheffield, Sheffield S1 3JD, UK, and also with the Department of Electrical Engineering, Princeton University, Princeton NJ 08540, USA.}
\thanks{Samir M. Perlaza is with the Institut National de Recherche en Informatique et Automatique (INRIA), Lyon, France, and also with the Department of Electrical Engineering, Princeton University, Princeton NJ 08540, USA.}
\thanks{Luis F. Ochoa is with the Department of Electrical and Electronic Engineering, The University of Melbourne, Melbourne 3010, Australia, and also with the School of Electrical and Electronic Engineering, The University of Manchester, Manchester M13 9PL, UK. (cgenes1@sheffield.ac.uk, esnaola@sheffield.ac.uk, samir.perlaza@inria.fr, luis$\char`_$ochoa@ieee.org, and d.coca@sheffield.ac.uk).}
}


%
\setlength\unitlength{1mm}

\newcommand{\insertfig}[3]{
\begin{figure}[htbp]\begin{center}\begin{picture}(120,90)
\put(0,-5){\includegraphics[width=12cm,height=9cm,clip=]{#1.eps}}\end{picture}\end{center}
\caption{#2}\label{#3}\end{figure}}

\newcommand{\insertxfig}[4]{
\begin{figure}[htbp]
\begin{center}
\leavevmode \centerline{\resizebox{#4\textwidth}{!}{\input
#1.pstex_t}}
\caption{#2} \label{#3}
\end{center}
\end{figure}}

\long\def\comment#1{}


\newfont{\bbb}{msbm10 scaled 700}
\newcommand{\CCC}{\mbox{\bbb C}}

\newfont{\bb}{msbm10 scaled 1100}
\newcommand{\CC}{\mbox{\bb C}}
\renewcommand{\SS}{\mbox{\bb S}}
\newcommand{\RR}{\mbox{\bb R}}
\newcommand{\PP}{\mbox{\bb P}}
\newcommand{\QQ}{\mbox{\bb Q}}
\newcommand{\ZZ}{\mbox{\bb Z}}
\newcommand{\FF}{\mbox{\bb F}}
\newcommand{\GG}{\mbox{\bb G}}
\newcommand{\EE}{\mbox{\bb E}}
\newcommand{\NN}{\mbox{\bb N}}
\newcommand{\KK}{\mbox{\bb K}}


\newcommand{\av}{{\bf a}}
\newcommand{\bv}{{\bf b}}
\newcommand{\cv}{{\bf c}}
\newcommand{\dv}{{\bf d}}
\newcommand{\ev}{{\bf e}}
\newcommand{\fv}{{\bf f}}
\newcommand{\gv}{{\bf g}}
\newcommand{\hv}{{\bf h}}
\newcommand{\iv}{{\bf i}}
\newcommand{\jv}{{\bf j}}
\newcommand{\kv}{{\bf k}}
\newcommand{\lv}{{\bf l}}
\newcommand{\mv}{{\bf m}}
\newcommand{\nv}{{\bf n}}
\newcommand{\ov}{{\bf o}}
\newcommand{\pv}{{\bf p}}
\newcommand{\qv}{{\bf q}}
\newcommand{\rv}{{\bf r}}
\newcommand{\sv}{{\bf s}}
\newcommand{\tv}{{\bf t}}
\newcommand{\uv}{{\bf u}}
\newcommand{\wv}{{\bf w}}
\newcommand{\vv}{{\bf v}}
\newcommand{\xv}{{\bf x}}
\newcommand{\yv}{{\bf y}}
\newcommand{\zv}{{\bf z}}
\newcommand{\ellv}{{\bf \ell}}
\newcommand{\zerov}{{\bf 0}}
\newcommand{\onev}{{\bf 1}}


\newcommand{\Am}{{\bf A}}
\newcommand{\Bm}{{\bf B}}
\newcommand{\Cm}{{\bf C}}
\newcommand{\Dm}{{\bf D}}
\newcommand{\Em}{{\bf E}}
\newcommand{\Fm}{{\bf F}}
\newcommand{\Gm}{{\bf G}}
\newcommand{\Hm}{{\bf H}}
\newcommand{\Id}{{\bf I}}
\newcommand{\Jm}{{\bf J}}
\newcommand{\Km}{{\bf K}}
\newcommand{\Lm}{{\bf L}}
\newcommand{\Mm}{{\bf M}}
\newcommand{\Nm}{{\bf N}}
\newcommand{\Om}{{\bf O}}
\newcommand{\Pm}{{\bf P}}
\newcommand{\Qm}{{\bf Q}}
\newcommand{\Rm}{{\bf R}}
\newcommand{\Sm}{{\bf S}}
\newcommand{\Tm}{{\bf T}}
\newcommand{\Um}{{\bf U}}
\newcommand{\Wm}{{\bf W}}
\newcommand{\Vm}{{\bf V}}
\newcommand{\Xm}{{\bf X}}
\newcommand{\Ym}{{\bf Y}}
\newcommand{\Zm}{{\bf Z}}


\newcommand{\Ac}{{\cal A}}
\newcommand{\Bc}{{\cal B}}
\newcommand{\Cc}{{\cal C}}
\newcommand{\Dc}{{\cal D}}
\newcommand{\Ec}{{\cal E}}
\newcommand{\Fc}{{\cal F}}
\newcommand{\Gc}{{\cal G}}
\newcommand{\Hc}{{\cal H}}
\newcommand{\Ic}{{\cal I}}
\newcommand{\Jc}{{\cal J}}
\newcommand{\Kc}{{\cal K}}
\newcommand{\Lc}{{\cal L}}
\newcommand{\Mc}{{\cal M}}
\newcommand{\Nc}{{\cal N}}
\newcommand{\Oc}{{\cal O}}
\newcommand{\Pc}{{\cal P}}
\newcommand{\Qc}{{\cal Q}}
\newcommand{\Rc}{{\cal R}}
\newcommand{\Sc}{{\cal S}}
\newcommand{\Tc}{{\cal T}}
\newcommand{\Uc}{{\cal U}}
\newcommand{\Wc}{{\cal W}}
\newcommand{\Vc}{{\cal V}}
\newcommand{\Xc}{{\cal X}}
\newcommand{\Yc}{{\cal Y}}
\newcommand{\Zc}{{\cal Z}}


\newcommand{\alphav}{\hbox{\boldmath$\alpha$}}
\newcommand{\betav}{\hbox{\boldmath$\beta$}}
\newcommand{\gammav}{\hbox{\boldmath$\gamma$}}
\newcommand{\deltav}{\hbox{\boldmath$\delta$}}
\newcommand{\etav}{\hbox{\boldmath$\eta$}}
\newcommand{\lambdav}{\hbox{\boldmath$\lambda$}}
\newcommand{\epsilonv}{\hbox{\boldmath$\epsilon$}}
\newcommand{\nuv}{\hbox{\boldmath$\nu$}}
\newcommand{\muv}{\hbox{\boldmath$\mu$}}
\newcommand{\zetav}{\hbox{\boldmath$\zeta$}}
\newcommand{\phiv}{\hbox{\boldmath$\phi$}}
\newcommand{\psiv}{\hbox{\boldmath$\psi$}}
\newcommand{\thetav}{\hbox{\boldmath$\theta$}}
\newcommand{\tauv}{\hbox{\boldmath$\tau$}}
\newcommand{\omegav}{\hbox{\boldmath$\omega$}}
\newcommand{\xiv}{\hbox{\boldmath$\xi$}}
\newcommand{\sigmav}{\hbox{\boldmath$\sigma$}}
\newcommand{\piv}{\hbox{\boldmath$\pi$}}
\newcommand{\rhov}{\hbox{\boldmath$\rho$}}

\newcommand{\Gammam}{\hbox{\boldmath$\Gamma$}}
\newcommand{\Lambdam}{\hbox{\boldmath$\Lambda$}}
\newcommand{\Deltam}{\hbox{\boldmath$\Delta$}}
\newcommand{\Sigmam}{\hbox{\boldmath$\Sigma$}}
\newcommand{\Phim}{\hbox{\boldmath$\Phi$}}
\newcommand{\Pim}{\hbox{\boldmath$\Pi$}}
\newcommand{\Psim}{\hbox{\boldmath$\Psi$}}
\newcommand{\Thetam}{\hbox{\boldmath$\Theta$}}
\newcommand{\Omegam}{\hbox{\boldmath$\Omega$}}
\newcommand{\Xim}{\hbox{\boldmath$\Xi$}}

\newcommand{\supp}{{\hbox{supp}}}
\newcommand{\sinc}{{\hbox{sinc}}}
\newcommand{\diag}{{\hbox{diag}}}
\renewcommand{\det}{{\hbox{det}}}
\newcommand{\trace}{{\hbox{tr}}}
\newcommand{\sign}{{\hbox{sign}}}
\renewcommand{\arg}{{\hbox{arg}}}
\newcommand{\var}{{\hbox{var}}}
\newcommand{\cov}{{\hbox{cov}}}
\newcommand{\SINR}{{\sf SINR}}
\newcommand{\SNR}{{\sf SNR}}
\newcommand{\Ei}{{\rm E}_{\rm i}}
\renewcommand{\Re}{{\rm Re}}
\renewcommand{\Im}{{\rm Im}}
\newcommand{\eqdef}{\stackrel{\Delta}{=}}
\newcommand{\defines}{{\,\,\stackrel{\scriptscriptstyle \bigtriangleup}{=}\,\,}}
\newcommand{\<}{\left\langle}
\renewcommand{\>}{\right\rangle}
\newcommand{\herm}{{\sf H}}
\newcommand{\transp}{{\sf T}}
\renewcommand{\vec}{{\rm vec}}


\newcommand{\GameNF}{\mathcal{G} = \left(\mathcal{K}, \left\lbrace\mathcal{A}_k \right\rbrace_{k \in \mathcal{K}},\phi \right)}
\newcommand{\gameNF}{\mathcal{G}}
\newcommand{\BR}{\mathrm{BR}}

\newenvironment{claim}[1]{\par\noindent\underline{Claim:}\space#1}{}
\newenvironment{claimproof}[1]{\par\noindent\underline{Proof:}\space#1}{\hfill $\blacksquare$}

\maketitle
\IEEEdisplaynontitleabstractindextext
\IEEEpeerreviewmaketitle

\begin{abstract}
The advanced operation of future electricity distribution systems is likely to require significant observability of the different parameters of interest (e.g., demand, voltages, currents, etc.). Ensuring completeness of data is, therefore, paramount. In this context, an algorithm for recovering missing state variable observations in electricity distribution systems is presented. The proposed method exploits the low rank structure of the state variables via a matrix completion approach while incorporating prior knowledge in the form of second order statistics. Specifically, the recovery method combines nuclear norm minimization with Bayesian estimation. The performance of the new algorithm is compared to the information-theoretic limits and tested trough simulations using real data of an urban low voltage distribution system. The impact of the prior knowledge is analyzed when a mismatched covariance is used and for a Markovian sampling that introduces structure in the observation pattern. Numerical results demonstrate that the proposed algorithm is robust and outperforms existing state of the art algorithms.
\end{abstract}

\vspace{-3mm}
\section{Introduction}
\IEEEPARstart{T}{he} wide-spread adoption of residential scale low carbon technologies, from PV systems to electric vehicles, will undoubtedly bring technical challenges to the electricity distribution systems as they have been designed for passive loads \cite{NO_16} and \cite{WSDBK_08}. Therefore, and as part of the Smart Grid vision, electricity distribution systems, including low voltage circuits, are likely to adopt a more active role so as to cost-effectively manage controllable network elements as well as participants\cite{LO_16}.  As a result, monitoring and control procedures are expected to face increasingly demanding performance requirements posed by the dynamic and unknown scenarios that the smart grid gives rise to.  Advanced control strategies require timely and accurate data describing the state of the grid. In this setting, the sensing infrastructure is expected to provide complete and reliable state information of the distribution system. However, in practical scenarios, the operator faces challenges like data injection attacks\cite{oevkp13},\cite{KJTT10} or missing data \cite{GWGCFS16}, \cite{GEPOC16}. Sensor failures, unreliable communication or data storage issues are some of the causes for incomplete sets of observations. As a consequence, the state of the grid is not perfectly known and control mechanisms are difficult to implement. For instance, accurate measurements are necessary to implement a centralized control scheme for voltage regulation in distribution systems\cite{IYFIOOH_16}. In view of this, it is vital to develop estimation procedures for the missing data using the available observations.

Missing data recovery can be cast as a minimum mean square error (MMSE) estimation problem when a probabilistic description of the underlying process governing the state variables is available. However, the MMSE estimation relies on accurate second order statistics which is an unrealistic assumption in practical scenarios \cite{ETG13,GEPOC16}. The increased number of nonlinear loads and the turbulent nature of distributed generation options in the locally controlled grid affects the precision of the postulated statistics for the state variables. For that reason, the efficiency of MMSE estimation is limited in the smart grid context\cite{GEPOC16}.

Matrix completion (MC) was recently proposed to recover missing data from partial observations \cite{CR12}. The main advantage is that the recovery via MC requires mild assumptions about the setting, e.g. access to second order statistics is not required. Instead, matrix completion-based recovery exploits the fact that correlated state variable vectors give rise to approximately low rank data matrices. That being the case, the recovery of the missing entries of low rank matrices is feasible in a convex optimization context provided that a sufficient fraction of the entries is observed \cite{CR12}, \cite{CT10}, and \cite{CP10}. The key theoretical results therein are based on the assumption that the locations of sampled entries are uniformly distributed. In practice, however, this assumption is not always satisfied. For instance, in electricity distribution systems missing data entries tend to display significant structure across both space and time \cite{GEPOC16}. The applicability of MC recovery for non-uniform sampling is studied in \cite{GWGCFS16}, \cite{MJI09}. Not surprisingly, low rank minimization tools are also used to address the problem of electricity price forecasting \cite{KZG14} and to develop a framework for efficient processing of synchrophasor data \cite{GWGCFS_16}.

This paper compares the performance of different missing data recovery strategies with respect to the information-theoretic limit and introduces a novel algorithm that addresses the main shortcomings of existing techniques. The performance of the new algorithm is tested against singular value thresholding (SVT) recovery\cite{CCS10} and MMSE estimation under realistic assumptions, i.e., the postulated statistics are not accurate and the sampling pattern is not uniform. In particular, a mismatched covariance matrix model is used in \cite{ETP12} and \cite{Verdu10} for the case in which imperfect second order statistics are available. Non-uniform sampling is considered to account for structure on the observation pattern\cite{GWGCFS_16}. Numerical results show a significant gain in performance for both cases when compared to SVT recovery. Remarkably, the new algorithm is robust to mismatched statistics and to non-uniform sampling patterns.
\vspace{-1.5mm}
\section{System model}

Consider a low voltage distribution system with $L$ feeders. Each feeder includes a sensing unit that measures the electrical magnitudes of operational interest at predetermined time instants.
These measurements that include phase active power, phase reactive power and phase voltage support the operator in controlling, monitoring, and managing the network. In practice, the acquisition process provides the operator with a noisy and incomplete set of state variables. For that reason, the operator needs to recover the missing data using the available observations.
\vspace{-1.5mm}
\subsection{Source Model for State Variables} \label{sec:source_model}
For a given electrical magnitude $s$, let $m_{i,j}^{(s)}$ be the value of that particular electrical magnitude at feeder $i$ at time $j$. 
The matrix of state variables for $s$, denoted by $\Mm^{(s)}\in \mathbb{R}^{N \times L}$, contains the aggregated state variable vectors from all feeders, i.e. $\Mm^{(s)}\eqdef[\mv_{1}^{(s)}, \mv_{2}^{(s)}, ...,\mv_{L}^{(s)}]$. The state variable vector for $s$, in feeder $i$, contains the $N$ state variables generated in the feeder and is given by $\mv_{i}^{(s)}=[m_{i,1}^{(s)}, m_{i,2}^{(s)}, ...,m_{i,N}^{(s)}]^T \in \mathbb{R}^{N}$. Without loss of generality the analysis is carried out for a particular electrical magnitude, and therefore, the index $s$ is dropped. The resulting data matrix $\Mm$ contains the state variable of interest at time instants $1,2,...,N$ for all $L$ feeders. 

Real data is used to model the statistical structure of the data generated in an electricity distribution system. The real data set under consideration contains values from 200 residential secondary substations across the North West of England collected from June 2013 to January 2014. The data collection is part of the  $\text{``Low Voltage Network Solutions"}$ project run by Electricity North West Limited \cite{enwl}. Each substation creates a daily file containing values of voltage, current and power levels for all three phases. An analysis of the distribution and sample covariance matrix of the voltage data set under consideration is presented in \cite{GEPOC16}. Therein it is shown that state variables can be modelled as a multivariate Gaussian random process
\begin{equation} \label{eq:Gmodel}
\mv_i {\sim} \Nc(\muv,\mathbf{\Sigma}),
\end{equation}
and $\lbrace\mv_i\rbrace_{i=1}^L$ is an independent and identically distributed sequence of random variables. Consequently, $\Mm$ is a realization of the random process describing the value of the state variable of interest across the grid.

\begin{figure}[t]
\centering
\includegraphics[width=3.5in]{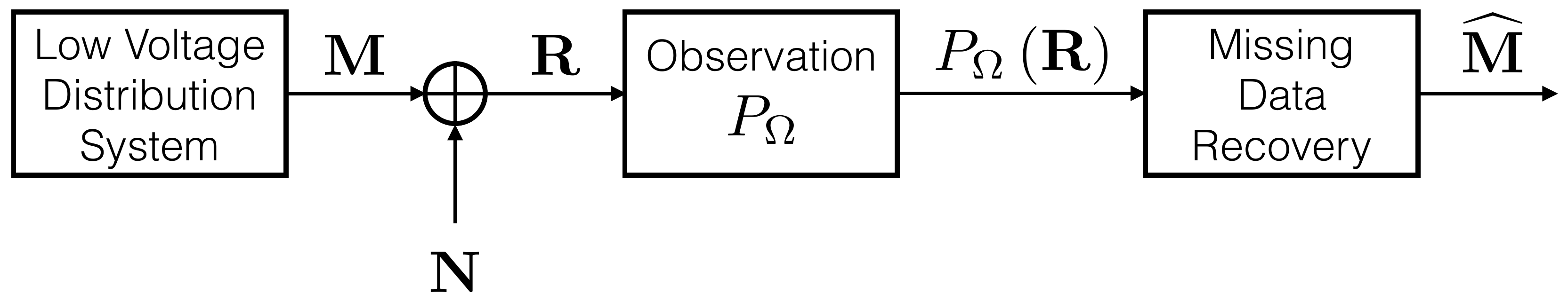}
\caption{Block diagram describing the system model.}
\label{fig:block}
\vspace{-5mm}
\end{figure}

Fig. \ref{fig:block} describes the distribution system monitoring model. In this setting, the electrical magnitudes describing the state of the system are modelled as a random process that outputs a realization $\Mm \in \mathbb{R}^{N \times L}$ every $N$ time instants. The state of the grid is fully described by the entries of the matrix $\Mm$.
However, the operator observes a subset of the complete set of state variables, i.e. measurements are lost during the acquisition process. The aim of the estimation process is to recover the missing entries.
\vspace{-1.5mm}
\subsection{Acquisition}
The sensing infrastructure introduces additive white Gaussian noise (AWGN) as a result of the thermal noise present at each sensor. The resulting measurements are given by
\begin{equation}
\Rm=\Mm+\Nm,
\end{equation}
where 
\begin{equation} \label{eq:noise}
(\Nm)_{i,j} \sim \Nc(0,\sigma^2),
\end{equation}
for $i \in \{1,2,...,N\}$ and  $j \in \{1,2,...,L\}$. Moreover, it is also assumed that only a fraction of the complete set of measurements (entries in $\Rm$) are communicated to the operator. Denote by $\Omega$ the subset of observed entries, i.e., $\Omega \subseteq \{1,2,...N\} \times \{1,2.,...,L\}$. By definition it follows that $\Omega$ is given by
\begin{equation}
\Omega \eqdef \{(i,j) : (\Rm)_{i,j} \textnormal{ is observed}\}.
\end{equation}
Formally, the acquisition process is modelled by the function $f :  \mathbb{R}^{N \times L} \to  \mathbb{R}^{|\Omega|}$ with $f(\Mm)=P_{\Omega}(\Rm)$ where
\begin{equation} \label{eq:obs}
P_{\Omega}(\Rm)=(\Rm)_{\Omega},
\end{equation}
and $|\Omega|$ denotes the cardinality of $\Omega$.
The observations given by (\ref{eq:obs}) describe all the data that is available to the operator for estimation purposes and therefore, the recovery of the missing data is performed from the observations $P_{\Omega}(\Rm)$.
\vspace{-3mm}
\subsection{Estimation}
The estimation process of the complete matrix of state variables based on the available observations is modelled by the function $g:\mathbb{R}^{|\Omega|} \to \mathbb{R}^{N \times L}$. The estimate $\widehat{\Mm}=g(f(\Mm))$ is obtained by solving an optimization problem based on an optimality criterion. In this paper, the optimality criterion is the mean square error (MSE) given by
\begin{equation}
\textnormal{MSE} \left( \Mm;g \right)= \frac{\mathbb{E}\left[ \|\Mm-g(f(\Mm))\|^2_F\right ]}{NL},
\end{equation}
where $\|\cdot\|_F$ denotes the Frobenius norm. The normalized mean square error (NMSE) is defined as
\begin{equation} \label{mse_to_rmse}
\textnormal{NMSE} \left( \Mm;g \right) = \textnormal{MSE} \left( \Mm;g \right) \frac{NL}{\|\Mm\|_F^2}.
\end{equation}
For this optimality criterion, the optimal estimate of the missing data is given by the MMSE estimate
\begin{equation}
\widehat{\Mm}_{\textnormal{MMSE}}=\mathbb{E}[\Mm|f(\Mm),\Sigmam],
\end{equation}
where $\Sigmam \in \mathbb{R}^{N \times N}$ is the covariance matrix defined in (\ref{eq:Gmodel}).
Note that, in general, obtaining the optimal estimate $\widehat{\Mm}_{\textnormal{MMSE}}$ requires knowledge of the probability distribution describing the state variables. If the state variables follow a Gaussian distribution it boils down to knowledge of the second order moments, i.e. the covariance matrix $\Sigmam$ which needs to be known prior to the estimation process. In practice, the operator relies on postulated statistics that typically do not match the actual statistics. Consequently, the accuracy of the estimate is a function of the difference between the real and the postulated statistics.
       
 \section{Information-Theoretic Limit} \label{sec:opta}
In order to assess the performance of the missing data recovery techniques in absolute terms, this section introduces the optimal performance theoretically attainable (OPTA) by an estimator $g$ when the state variables follow a multivariate Gaussian distribution. For a given number of observations, the minimum distortion achievable by any estimation method is determined by the rate-distortion function \cite{cover_12}. In the electricity distribution setting described above, the observations entries are corrupted by additive white Gaussian noise which determines the finite rate at which information about the state variables is obtained from the observations. Consequently, the optimal performance is bounded by the capacity of the AWGN channel
\begin{equation}
R({D})<C,
\end{equation}
where $R$ is the rate at which the source needs to be observed to achieve a distortion $D$, and $C$ is the capacity of the parallel AWGN channels modelling the observation process.
In view of this, the OPTA for a multivariate Gaussian source is given by
\begin{equation}
R({D}) \leq \frac{|\Omega|}{2NL} \textnormal{log}_{10}(1+ {\sf snr}),
\end{equation}
where the signal to noise ratio, denoted by ${\sf snr}$, is defined as
\begin{equation}
{\sf snr}\eqdef\frac{\frac{1}{N}\textnormal{Tr}(\Sigmam)}{\sigma^2},
\end{equation}
where $\sigma^2$ is defined in (\ref{eq:noise}).
The rate-distortion function of a multivariate Gaussian process is computed using the following parametric equations \cite{Kolmogorov56} 
\begin{equation}
\label{eq:RD}
\begin{cases}
R(\theta) & = \frac{1}{N} \sum^{N-1}_{i=0} \textnormal{max}(0, \frac{1}{2} \log \frac{\lambda_i}{\theta}) \\
D(\theta) & = \frac{1}{N} \sum^{N-1}_{i=0} \textnormal{min}(\theta, \lambda_i),
\end{cases}
\end{equation}
where $R$ is the source rate in nats/symbol, $D$ is the mean square error distortion per entry, $\lambda_i$ is the $i \,$th largest eigenvalue of $\mathbf{\Sigma}$, and $\theta$ is a parameter. The NMSE theoretically attainable, $\textnormal{NMSE}(\Mm;\textnormal{OPTA})$, follows from combining (\ref{mse_to_rmse}) and (\ref{eq:RD}) and is determined by
\begin{equation}
\textnormal{NMSE}(\Mm;\textnormal{OPTA})={D} \frac{NL}{\|\Mm\|_F^2} .
\end{equation}  
\section{Recovery of missing data}
In this section, the information-theoretic limit for missing data recovery presented in Section \ref{sec:opta}, is compared with MMSE estimation and the singular value thresholding (SVT) recovery.
\vspace{-1.5mm}
\subsection{Minimum Mean Squared Error Estimation}
Linear MMSE (LMMSE) estimation achieves the optimal performance in the recovery of missing data for a given set of observations $\Omega$ when the data is generated by a multivariate Gaussian source and the optimality criteria is the MSE. However, this estimation procedure relies on access to second order statistics of the state variables.
In particular, each column of the matrix $P_{\Omega}(\Rm)$ is given by
\begin{equation}
P_{\Omega}(\rv_i)=\Am_i (\mv_i + \nv_i),
\end{equation}
where $\Am_i$ is defined such that $\Am_i \mv_i= P_{\Omega}(\mv_i)$ and $i \in \{1,2,...,L\}$.
Consequently, the LMMSE estimate for each state variable vector is given by
\begin{equation} \label{d_mmse_def}
\widehat{\mv}_{i}=\muv + \Gammam_i (P_{\Omega}(\rv_i)- \Am_i \muv),
\end{equation}
where $\muv$ is defined in (\ref{eq:Gmodel}) and
\begin{equation} 
\Gammam_i = \Sigmam \Am_i^T (\Am_i \Sigmam \Am_i^T + \sigma^2 \mathbf{I})^{-1},
\end{equation}
$i \in \{1,2,...,L\}$.
The normalized error achieved by the LMMSE estimator is given by        
\begin{equation}
\textnormal{NMSE}(\Mm;\textnormal{LMMSE}) = \frac{\|\Mm-\widehat{\Mm}_{\textnormal{LMMSE}}\|_F^2}{\|\Mm\|_F^2},
\end{equation}
where $\widehat{\Mm}_{\textnormal{LMMSE}}=[\widehat{\mv}_{1}, \widehat{\mv}_{2}, ..., \widehat{\mv}_{L}]$, with $\widehat{\mv}_{i}$ defined in (\ref{d_mmse_def}).
\vspace{-1.5mm}
\subsection{Singular Value Thresholding}
Low rank matrices are recovered from a subset of the entries via rank minimization techniques under mild coherence conditions on the set of observations \cite{CR12}. Specifically, the missing entries are recovered by solving the following rank minimization problem:
\begin{equation} \label{nonconvex}
\begin{aligned}
& {\underset{\mathbf{X}}{\text{minimize}}}
& & \mathrm{rank}(\mathbf{X}) \\
& \text{subject to}
& & P_{\Omega}(\mathbf{X})=P_{\Omega}(\mathbf{M}).\\
\end{aligned}
\end{equation}
Unfortunately, this rank minimization problem is NP-hard.
Favorably, in \cite{CR12} it is shown that when the entries on $\Omega$ are sampled uniformly at random, the solution of the rank minimization problem in (\ref{nonconvex}) is obtained with high probability by solving the nuclear norm minimization problem in (\ref{convex}).

SVT is a matrix completion algorithm \cite{CCS10} which produces a sequence of matrices $\Xm^k$ that converges to the unique solution of the following optimization problem:
\begin{equation} \label{svt}
\begin{aligned}
& {\underset{\mathbf{X}}{\text{minimize}}}
& & \tau \mathrm  \lVert \mathbf{X} \rVert_{*}+ \frac{1}{2}\lVert \mathbf{X} \rVert_{F}^{2} \\
& \text{subject to}
& & P_{\Omega}(\mathbf{X})=P_{\Omega}(\mathbf{M}),\\
\end{aligned}
\end{equation}
 where $ \lVert \mathbf{X} \rVert_{*}$ is the nuclear norm of the matrix $\mathbf{X}$. Note that when $\tau \to \infty$, the optimization problem in (\ref{svt}) converges to the nuclear norm minimization problem proposed in \cite{CR12}
\begin{equation} \label{convex}
\begin{aligned}
& {\underset{\mathbf{X}}{\text{minimize}}}
& & \mathrm \lVert \mathbf{X} \rVert_{*} \\
& \text{subject to}
& & P_{\Omega}(\mathbf{X})=P_{\Omega}(\mathbf{M}).\\
\end{aligned}
\end{equation}

For large values of $\tau$, SVT provides the solution to the nuclear norm minimization problem.
Compared to alternatives like SeDuMi \cite{strum99} or SDPT3 \cite{TTT99}, SVT features a lower computational cost per iteration. This is achieved by exploiting the sparsity of $\Ym^k$ and the low-rank property of $\Xm^k$ to reduce storage requirements.
The low computational cost results in the possibility of using larger matrices. Simulation results in \cite{CCS10} show that SVT recovers matrices with nearly a billion entries. In comparison, SeDuMi and SDPT3 produce accurate solutions for squared matrices with dimension close to fifty. In \cite{LV09} the structure of the problem is exploited to reduce the memory requirements and increase the matrix size up to 350. Because of the dimension of the data sets produced by low voltage distribution systems, the remaining of the paper focuses on the SVT as a benchmark MC-based recovery.
The main idea in SVT consists in the following iteration steps:
\begin{equation}
\begin{cases}
\mathbf{X}^k= D_{\tau}(\mathbf{Y}^{k-1}), \\
\mathbf{Y}^k=\mathbf{Y}^{k-1}+\delta_s P_{\Omega}(\mathbf{M}-\mathbf{X}^k),\\
\end {cases}
\end{equation}
where $\mathbf{Y}^0=\mathbf{0}$, $\delta_s$ is the step size that obeys $0 < \delta_s < 2$, and the soft-thresholding operator, $D_{\tau}$, applies a soft-thresholding rule to the singular values of $\mathbf{Y}^{k-1}$, shrinking these towards zero. For a matrix $\mathbf{Y} \in \mathbb{Re}^{N \times L}$ of rank $r$ with singular value decomposition given by
\begin{equation}
\mathbf{Y}=\mathbf{U} \mathbf{S} \mathbf{V}^T,  \quad
\mathbf{S}= \textnormal{diag}(\{\sigma_i(\Ym)\}_{1 \leq i \leq r}),
\end{equation}
where $\mathbf{U}$ and $\mathbf{V}$ are unitary matrices of size $N \times r$ and $L \times r$, respectively, and $\sigma_i(\Ym)$ are the singular values of the matrix $\mathbf{Y}$, the soft-thresholding operator is defined as
\newline
\begin{equation} \label{eq:d_tau}
D_{\tau}(\mathbf{Y})\eqdef \mathbf{U} D_{\tau}(\mathbf{S}) \mathbf{V}^T, \textnormal{ with }
D_{\tau}(\mathbf{S})=\diag(\{(\sigma_i(\Ym)- \tau)_+\}),
\end{equation}
where $t_+= \textnormal{max}(0,t)$. 
Interestingly, the choice of $\tau$ is important to guarantee a successful recovery, since large values guarantee a low-rank matrix estimate but for values larger than $\underset{i}{\textnormal{max}} \,( \sigma_i(\Ym))$ all the singular values vanish. In \cite{CCS10}, the proposed threshold is $\tau =5N$. However, simulation results presented in \cite{GEPOC16} show that $\tau =5N$ gives suboptimal performance when the number of missing entries is large. Unfortunately, finding the optimal threshold when the matrix is sparse is still an open problem. In general, the value of the threshold for soft-thresholding based recovery algorithms is obtained via numerical optimization in \cite{GEPOC16} and \cite{CP10}.
The same soft-thresholding operator, $D_{\tau}$, is used in a different framework for denoising \cite{CP10},  \cite{DG14}, and \cite{JFW16}. In this context, the performance of the denoiser, measured in MSE, is estimated using Stein\textquotesingle s unbiased risk estimate (SURE) \cite{stein81}. In \cite{CST13} a closed-form expression for the unbiased risk estimate is presented for the operator $D_{\tau}$.
\vspace{-1.5mm}
\subsection{Performance Evaluation with Real Data} \label{spawc_num_res}
This subsection presents a comparison between LMMSE and SVT, and the theoretical limit, OPTA, using real electricity distribution system data. The test matrix, $\Mm$, is a square matrix of size $500$, i.e. $N=L=500$, and contains voltage measurements covering the state of the grid for a period of 2 hours. Each column represents a different state variable vector that describes the grid on a different day and for a different feeder.
The entries in $\Omega$ are sampled uniformly at random with probability
\begin{equation}
\mathbb{P}[(i,j) \in \Omega]= \frac{1}{NL}\mathbb{E}[|\Omega|],
\end{equation}
and the performance of the SVT-based recovery is defined in terms of the NMSE given by
\begin{equation}
\textnormal{NMSE}(\Mm;\textnormal{SVT})=\frac{\|\mathbf{M-\widehat{M}_\textnormal{SVT}}\|^2_F}{\|\Mm\|_F^2},
\end{equation}
where $\widehat{\Mm}_\textnormal{SVT}$ is the SVT estimate of $\Mm$ based on $P_{\Omega}(\Rm)$.
\begin{figure}[t!]
\centering
\includegraphics[width=3.7in]{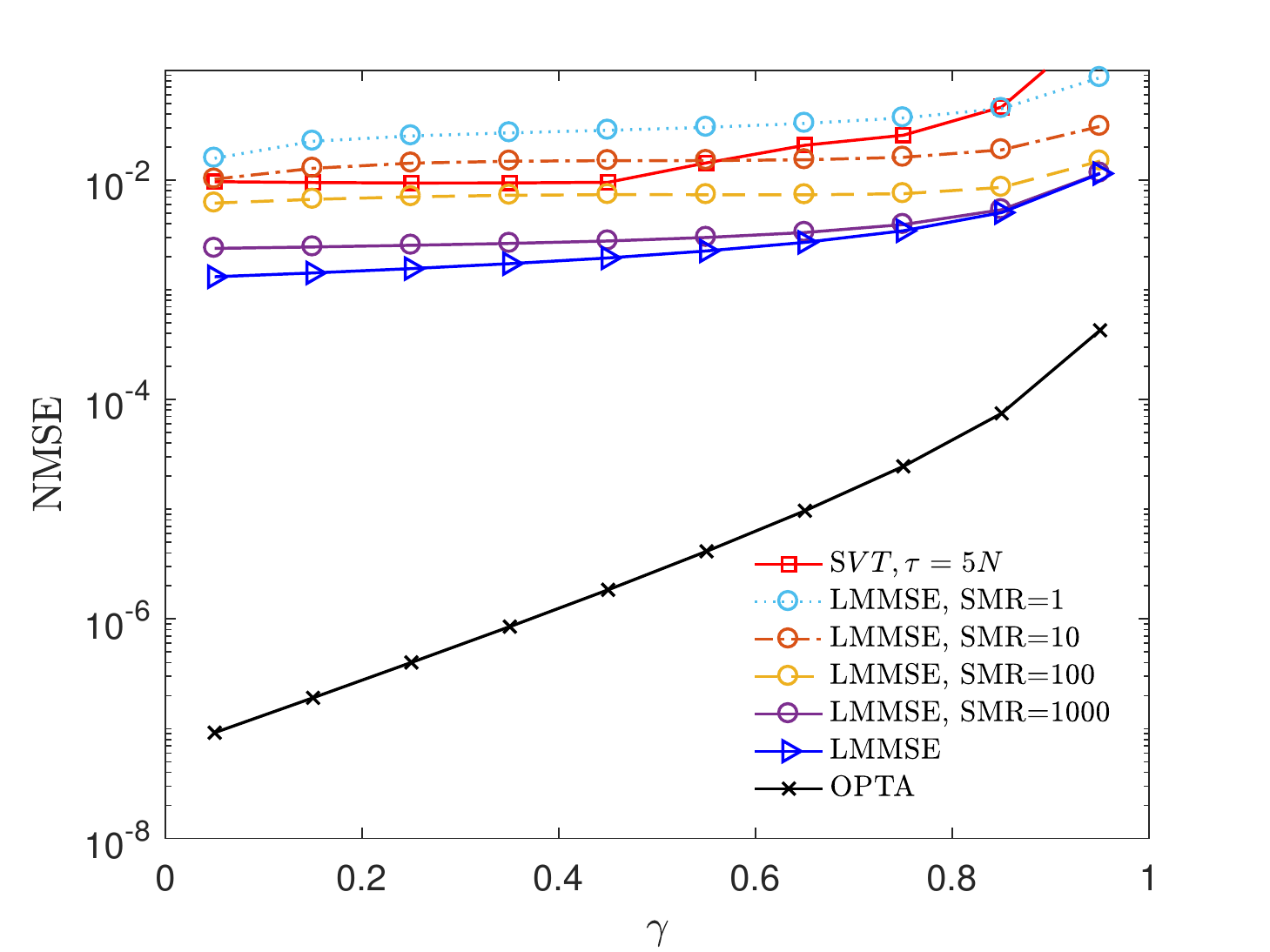}
\caption{Real data recovery performance using SVT, LMMSE estimation, for different levels of mismatch, and the OPTA, when  $\textnormal{SNR}=20$ dB.}
\label{fig:svt_mmse_rd}
\vspace{-5mm}
\end{figure}
Let $\gamma$ be the expected value of the ratio of missing entries for the matrix $\Mm$, that is:
\begin{equation} \label{eq:gamma}
\gamma \eqdef 1 - \frac{1}{NL}\mathbb{E}[|\Omega|].
\end{equation}
Since the performance of the LMMSE estimator depends on the covariance matrix $\Sigmam$, a mismatched covariance matrix model is introduced to account for the difference between the postulated and actual statistics. Specifically, the postulated covariance matrix is given by
\begin{equation} \label{eq:mismatch}
\mathbf{\Sigma}^*=\mathbf{\Sigma} + \frac{1}{\textnormal{SMR}} \frac{\|\Sigmam\|_F^2}{\|\mathbf{\Delta}\|_F^2} \mathbf{\Delta},
\end{equation}
 where $\mathbf{\Sigma}$ is the actual covariance matrix, $\mathbf{\Delta=H H}^T$ with $\mathbf{H} \in \mathbb{R}^{ N \times N}$, the entries of $\mathbf{H}$ are distributed as $\Nc(0,1)$.  The strength of the mismatch is determined by the signal to mismatch ratio (SMR), which is defined such that for $\textnormal{SMR}=1$ the norm of the mismatch is equal to the norm of the real covariance matrix, i.e., $\|\Sigmam\|_F^2=\|\alpha \mathbf{\Delta}\|_F^2$.
 
Fig. \ref{fig:svt_mmse_rd} shows the performance, measured in NMSE, for the SVT-based recovery compared to the performance of the LMMSE estimator when different levels of mismatch  are introduced and to the theoretical limit given by the OPTA. Numerical results in this section are obtained for a signal to noise ratio value of SNR$=20$ dB, where
$
\textnormal{SNR} \eqdef 10 \, \textnormal{log}_{10} {\sf snr}.
$
It can be seen that the performance of the SVT algorithm is closer to the theoretical limit when the number of missing entries is large. Interestingly, the LMMSE estimator gives better performance when $\textnormal{SMR}\geq100$. However, when SMR=10 and $\gamma \leq 0.55$ the SVT algorithm outperforms the LMMSE estimator. Moreover, the SVT provides a better recovery for SMR=1 for almost all values of $\gamma$. In view of this, the LMMSE estimation requires accurate second order statistics to perform competitively in this setting which is an unrealistic assumption in a practical scenario. Moreover, the performance of the SVT algorithm depends of the threshold $\tau$\cite{GEPOC16} which is difficult to optimize for this case.
\vspace{-2mm}
\section{Main Result}
This section introduces a novel algorithm for missing data recovery that incorporates imperfect second order information statistics. The new approach is based on the SVT algorithm but it exploits the information about the second order statistics to optimize the threshold $\tau$ at each iteration $k$.
\vspace{-2mm}
\subsection{Soft-thresholding parameter}
The main shortcoming of the SVT algorithm is the lack of guidelines for tuning the threshold $\tau$. Numerical results in \cite{GEPOC16} show that the value $5N$ proposed in \cite{CCS10} is not optimal for every scenario. In order to provide better recovery it is essential to tune the value of $\tau$ for each iteration of the algorithm. In SVT the soft-thresholding operator is applied on a sparse matrix which increases the difficulty of the tuning process.
\vspace{-2mm}
\subsection{Exploiting second order statistics}
In order to overcome the limitation imposed by the sparse structure of the matrix $\Ym^k$, the proposed algorithm estimates the missing entries prior to the soft-thresholding step. Thus, the available prior knowledge is exploited to produce an estimate of the entries not contained in $\Omega$. In this case, at each iteration $k$ of the proposed algorithm the matrix $\Zm^k$ is computed as
\begin{equation}
\Zm^k=\Ym^k+\Lm^k,
\end{equation}
where $\Ym^k$ is defined as in the SVT algorithm and $\Lm^k$ is the LMMSE estimate given by
\begin{equation}
\Lm^k=P_{\Omega^c}(\muv) + \Sigmam_{\Omega^c \Omega} \Sigmam_{\Omega \Omega}^{-1} (P_{\Omega}(\Ym^k)-P_{\Omega}(\muv)),
\end{equation}
where $\Omega$ is the set of observed entries, $\Omega^c$ is the set of missing entries, $\mathbf{\Sigma}_{\Omega^c \Omega}$ is the covariance matrix between the entries in $\Omega^c$ and the entries in $\Omega$ and  $\mathbf{\Sigma}_{\Omega \Omega}$ is the covariance matrix of the entries in $\Omega$.
In a nutshell, the unknown entries are estimated using the LMMSE-based recovery at each iteration $k$. The result is a complete matrix $\Zm^k$ for which the tuning of the threshold is feasible.
\vspace{-2mm}
\subsection{Optimization of thresholding parameter}
Using the main result in \cite{CST13}, the performance of the soft-thresholding operator can be estimated when the input matrix accepts the following model
\begin{equation} \label{eq:sure_model}
\Zm=\Mm+\Wm,
\end{equation}
where the entries of $\Wm$ are
\begin{equation}
(\Wm)_{i,j} \overset{iid}{\sim} \Nc(0,\sigma_{\Zm}^2),
\end{equation}
for $i \in \{1,2,...,N\}$ and $j \in \{1,2,...,L\}$.
In this setting, the SURE \cite{stein81} is given by
\begin{equation} \label{sure}
\begin{split}
\textnormal{SURE}(D_{\tau})(\Zm) = &-NL \sigma_{\Zm}^2 + \sum_{i=1}^{\textnormal{min}(N,L)} \textnormal{min}(\tau^2,\sigma_i^2(\Zm)) \\
&+ 2 \sigma_{\Zm}^2 \textnormal{div}(D_{\tau}(\Zm)),\\
\end{split}
\end{equation}
where $\sigma_i(\Zm)$ are is the $i$-th singular value of $\Zm$ for $i \in \{1,2,\ldots, N\}$. A closed-form expression for the divergence of this estimator is obtained in \cite{CST13}. For the case in which $\Zm\in\mathbb{R}^{N\times L}$ the divergence is given by
\begin{equation} \label{div}
\begin{split}
\textnormal{div}(D_{\tau}(\Zm)) = & \sum_{i=1}^{\textnormal{min}(N,L)} \bigg[\mathbb{I}(\sigma_i(\Zm)>\tau)+ |N-L|\frac{(\sigma_i(\Zm)-\tau)_+}{\sigma_i(\Zm)} \bigg] \\ 
& +2 \sum_{ i \neq j,i,j=1}^{\textnormal{min}(N,L)} \frac{\sigma_i(\Zm) (\sigma_i(\Zm)-\tau)_+}{\sigma_i^2(\Zm)-\sigma_j^2(\Zm)},
\end{split}
\end{equation}
when $\Zm$ has no repeated singular values and is zero otherwise.
Therefore, combining (\ref{sure}) and (\ref{div}) gives a closed-form expression for the performance of the soft-thresholding operator for different values of $\tau$ and different noise levels $\sigma_{\Zm}^2$.

The proposed algorithm approximates $\sigma_{\Zm}^2$ with the weighted sum of the noise in $\Omega$ and in $\Omega^c$. Consequently,  $\sigma_{\Zm^k}^2$ is calculated as
\begin{equation} \label{eq:sigma_zk}
\sigma_{\Zm^k}^2= \frac{\|\Ym^k-P_{\Omega}(\Mm)\|_F^2 + |\Omega^c| {D}_{\textnormal{LMMSE}}}{NL},
\end{equation}
where ${D}_{\textnormal{LMMSE}}$ represents the average noise per entry in $\Omega^c$.
The optimal threshold for the matrix $\Zm^k$ is denoted by $\tau_*^k$ and it is calculated using
\begin{equation}
\tau_*^k=\underset{\tau}{\textnormal{arg\,min}} \, \textnormal{SURE}(D_{\tau})(\Zm^k),
\end{equation}
where $\sigma_{\Zm^k}^2$ is given by (\ref{eq:sigma_zk}).
Therefore, the iterations of the proposed algorithm are
\begin{equation} \label{bsvt_def}
\begin{cases}
\mathbf{X}^k= D_{\tau}(\mathbf{Z}^{k-1}), \\
\mathbf{Y}^k=\mathbf{Y}^{k-1}+\delta_b P_{\Omega}(\mathbf{M}-\mathbf{X}^k),\\
\Zm^k=\Ym^k+\Lm^k,\\
\end {cases}
\end{equation}
where the $D_{\tau}$ is defined by (\ref{eq:d_tau}) and the step size $\delta_b$ is similar to the step size $\delta_s$ in the SVT algorithm. The initial conditions are $\Zm^0=\mathbf{0}$, $\Ym^0=\mathbf{0}$ and $\tau=0$. The stopping criteria is similar to the SVT algorithm, namely
\begin{equation}
\frac{\|P_{\Omega}(\Xm^k-\Mm)\|_F}{\|P_{\Omega}(\Mm)\|_F} \leq \epsilon.
\end{equation}
\begin{algorithm}
\caption{Bayesian Singular Value Thresholding}\label{alg:bsvt}
\begin{algorithmic}[1]
\Require observations set $\Omega$, and observed entries $P_{\Omega}(\Rm)$, mean $\muv$, covariance matrix $\Sigmam$, step size $\delta_b$, tolerance $\epsilon$, and maximum iteration count $k_{\textnormal{max}}$
\Ensure $\widehat{\Mm}_{\textnormal{BSVT}}$
\State Set $\Ym^0=\mathbf{0}$
\State Set $\Zm^0=\mathbf{0}$
\State Set $\tau=0$
\State Set $\Omega^c=\{1,2,...,N\} \times \{1,2,...,L\}\setminus \Omega$
\For {$k=1$ to $k_{\textnormal{max}}$}
\State Compute $[\Um, \Sm, \Vm]=\textnormal{svd}(\Zm^{k-1})$
\State Set $\Xm^k= \sum_{j=1}^{\textnormal{min}(N,L)} (\textnormal{max}(0,\sigma_j(\Zm^{k-1})-\tau)\uv_j \vv_j$
\If  {$\|P_{\Omega}(\Xm^k-\Mm)\|_F / \|P_{\Omega}(\Mm)\|_F \leq \epsilon$} {\bf break} \EndIf
\State Set $\Ym^k=\Ym^{k-1}+ \delta_b P_{\Omega}(\Mm-\Xm^k)$
\State Set $\Lm^k=P_{\Omega^c}(\muv) + \Sigmam_{\Omega^c \Omega} \Sigmam_{\Omega \Omega}^{-1} (\Ym^k-P_{\Omega}(\muv))$
\State Set $\Zm^k=\Ym^k+\Lm^k$
\State Set $\sigma_{\Zm^k}^2=(\|\Ym^k-P_{\Omega}(\Mm)\|_F^2 + |\Omega^c| {D}_{\textnormal{LMMSE}})  / NL$ 
\State Set $\tau=\underset{\tau}{\textnormal{arg\,min}} \, \textnormal{SURE}(D_{\tau})(\Zm^k)$ 
\EndFor
\State Set $\widehat{\Mm}_{\textnormal{BSVT}}= \Xm^k$
\end{algorithmic}
\end{algorithm}
The main advantage of the proposed algorithm is that the threshold is optimized at each iteration facilitated by the prior knowledge incorporated into the structure of the algorithm. First, an initial guess of the unavailable entries is formed, at each iteration $k$, based on $\Ym^k$ and the covariance matrix $\Sigmam$. The results are aggregated in the matrix $\Zm^k$ which is approximated by the model in (\ref{eq:sure_model}). In this case, an estimate of the noise level, $\sigma_{\Zm^k}^2$, is needed to compute the SURE. The optimal value of $\tau$ for $\Zm^k$ is obtained by minimizing $\textnormal{SURE}(D_{\tau})(\Zm^k)$ in (\ref{sure}).
Admittedly, the optimization of the threshold is only possible as long as second order statistics are available. Therefore, the new approach requires additional knowledge that is not necessary when using the SVT algorithm. That being said, the SVT algorithm requires setting the value for the threshold which in general is difficult to tune. The same amount of prior knowledge, i.e., covariance matrix, is required by the LMMSE estimator. Still, when the postulated statistics are not accurate, the performance of the LMMSE-based recovery reduces by up to an order of magnitude in NMSE (See Fig. \ref{fig:svt_mmse_rd}). For the proposed algorithm, the trade-off between the performance and the accuracy of the prior knowledge is studied in Section \ref{sec:numeric}.
\vspace{-1.5mm}
\section{Numerical Analysis} \label{sec:numeric}
This section analyzes the performance of the BSVT algorithm using the real data set presented in Section \ref{sec:source_model}. The data matrix $\Mm$, utilized to assess the performance of the proposed algorithm, is the same used in Section \ref{spawc_num_res} and contains the voltage measurements from the electricity distribution system. Similarly, the simulations in this section assume a signal to noise ratio value of SNR$=20$ dB. Moreover, the performance of the BSVT algorithm is also measured in terms of NMSE given by
\begin{equation}
\textnormal{NMSE}(\Mm;\textnormal{BSVT})=\frac{\| \Mm- \widehat{\Mm}_\textnormal{BSVT}\|^2_F}{\|\Mm\|_F^2},
\end{equation}
where $\widehat{\Mm}_\textnormal{BSVT}$ is the output of the BSVT recovery. The performance of each recovery technique is averaged over 20 realisations of $\Omega$ for each ratio of missing entries.
Numerically, the proposed algorithm is evaluated on three aspects. First, the gain in performance for the optimized threshold is assessed. The Section \ref{subsec:tau} compares the performance of the SVT-based recovery with the BSVT algorithm when accurate second order statistics are available. Secondly, the robustness of the BSVT recovery when perfect prior knowledge is not available is evaluated. A comparison between the SVT algorithm, the LMMSE estimator and the BSVT recovery is presented for different SMR values. The case in which perfect second-order statistics are available is also included. Finally, the robustness of the BSVT recovery to different sampling patterns is evaluated using Markov-chain-based sampling. The numerical performance of the new algorithm is compared to the SVT algorithm for the case in which the positions of the missing entries are not uniformly distributed.
\vspace{-1.5mm}
\subsection{Performance of the optimized threshold} \label{subsec:tau}
In this section, the performance of the new algorithm is compared to the SVT-based recovery using same data matrix $\Mm$ and the same sets of available entries, $\Omega$, for a particular ratio of missing entries $\gamma$ as defined in (\ref{eq:gamma}). The positions of the missing entries are sampled uniformly at random from the set of all entries.

\begin{figure}[t!]
\centering
\includegraphics[width=3.7in]{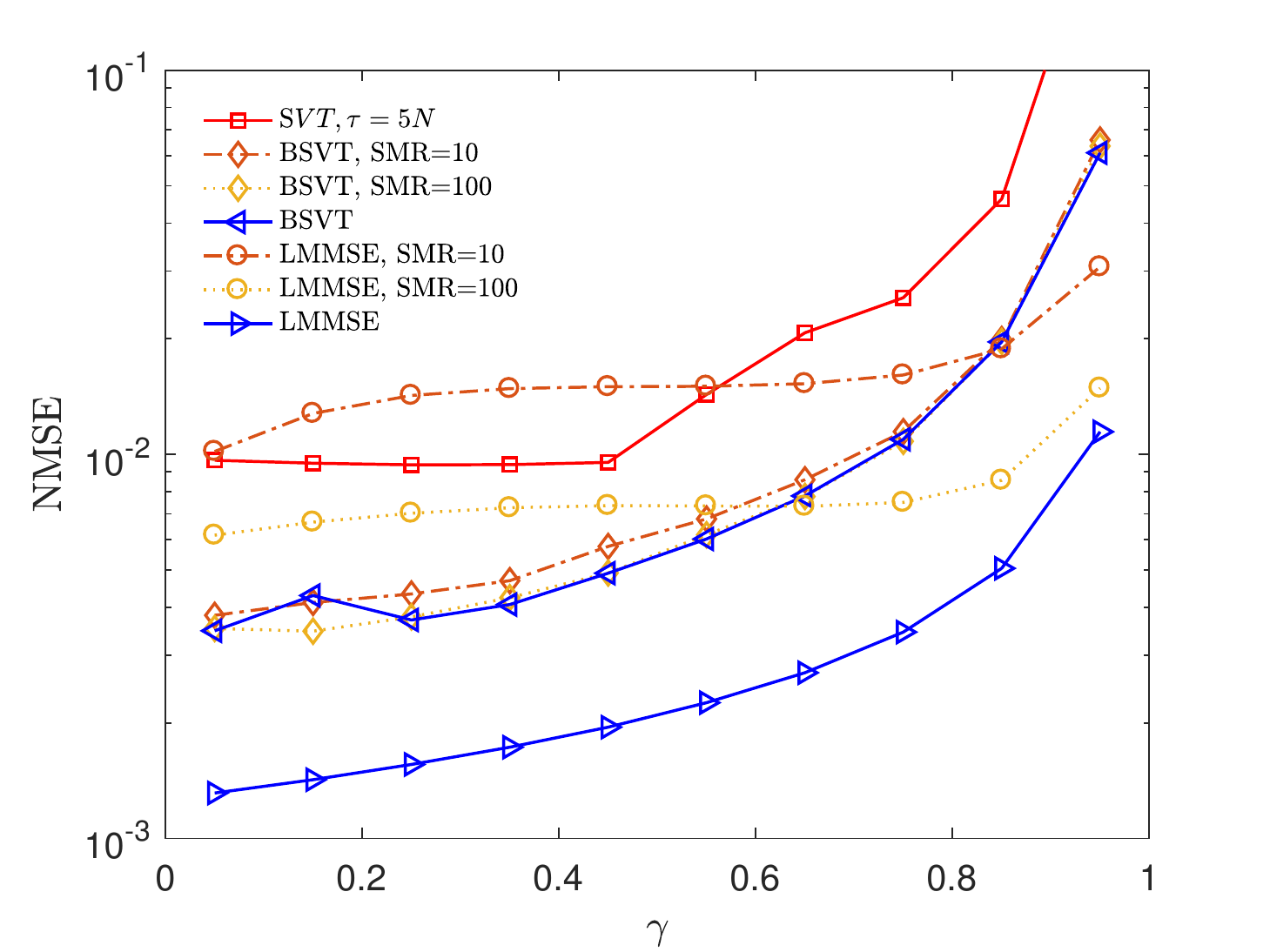}
\caption{Real data recovery performance using SVT, LMMSE estimation and BSVT for different levels of mismatch, when  $\textnormal{SNR}=20$ dB.}
\label{fig:svt_mmse_bsvt}
\vspace{-5mm}
\end{figure}

Fig. \ref{fig:svt_mmse_bsvt} depicts the performance of both algorithms when applied in identical scenarios. Clearly, the optimized threshold and the Bayesian estimation step increase the performance of the algorithm when accurate second order statistics are available. When the postulated statistics, i.e., those available to the operator are identical to the real statistics, the BSVT algorithm provides a better performance for all values of $\gamma$. The gain in performance is larger when the ratio of missing entries is smaller than $0.4$. Interestingly, the boost in performance is substantial in the region in which SVT is least efficient when compared to the fundamental limit (See Fig. \ref{fig:svt_mmse_rd}). However, in practical scenarios the postulated and actual statistics are different. The impact of mismatched statistics is considered in the following section.
\vspace{-1.5mm}
\subsection{Robustness with respect to mismatched statistics}
In order to address the problem of missing data recovery in a realistic scenario, a level of mismatch between the real covariance matrix and the one available to the operator is considered. The mismatch covariance matrix model presented in (\ref{eq:mismatch}) is used in this section to assess the sensibility of the proposed algorithm to inaccurate prior knowledge. Hence, the LMMSE estimator and the BSVT algorithm are compared in the no-mismatch regime and for a SMR value of 100 and 10. The performance of the SVT-based recovery is included as a benchmark for comparing rank minimization based approaches.

Fig. \ref{fig:svt_mmse_bsvt}  depicts the performance of the different estimation methods when mismatched second order statistics are available. Remarkably, the proposed algorithm is robust to mismatch in the second order statistics. In contrast with the LMMSE estimator, the performance of the BSVT algorithm does not change significantly when mismatch occurs. Moreover, the BSVT algorithm gives better recovery than the SVT-based recovery in all mismatch regimes throughout the range of $\gamma$. In comparison with the LMMSE estimation, the BSVT algorithm performs better for $\textnormal{SMR}=100$  when $\gamma\leq 0.65$. Furthermore, for  $\textnormal{SMR}=10$ the proposed approach is the best performing recovery method for almost all values of $\gamma$.
In practical scenarios, when the mismatch regime is difficult to establish, the choice between LMMSE and SVT is difficult to make. BSVT is a robust alternative and gives better recovery in a wide range of missing date regimes.

\begin{figure}[t!]
\centering
\includegraphics[width=3.7in]{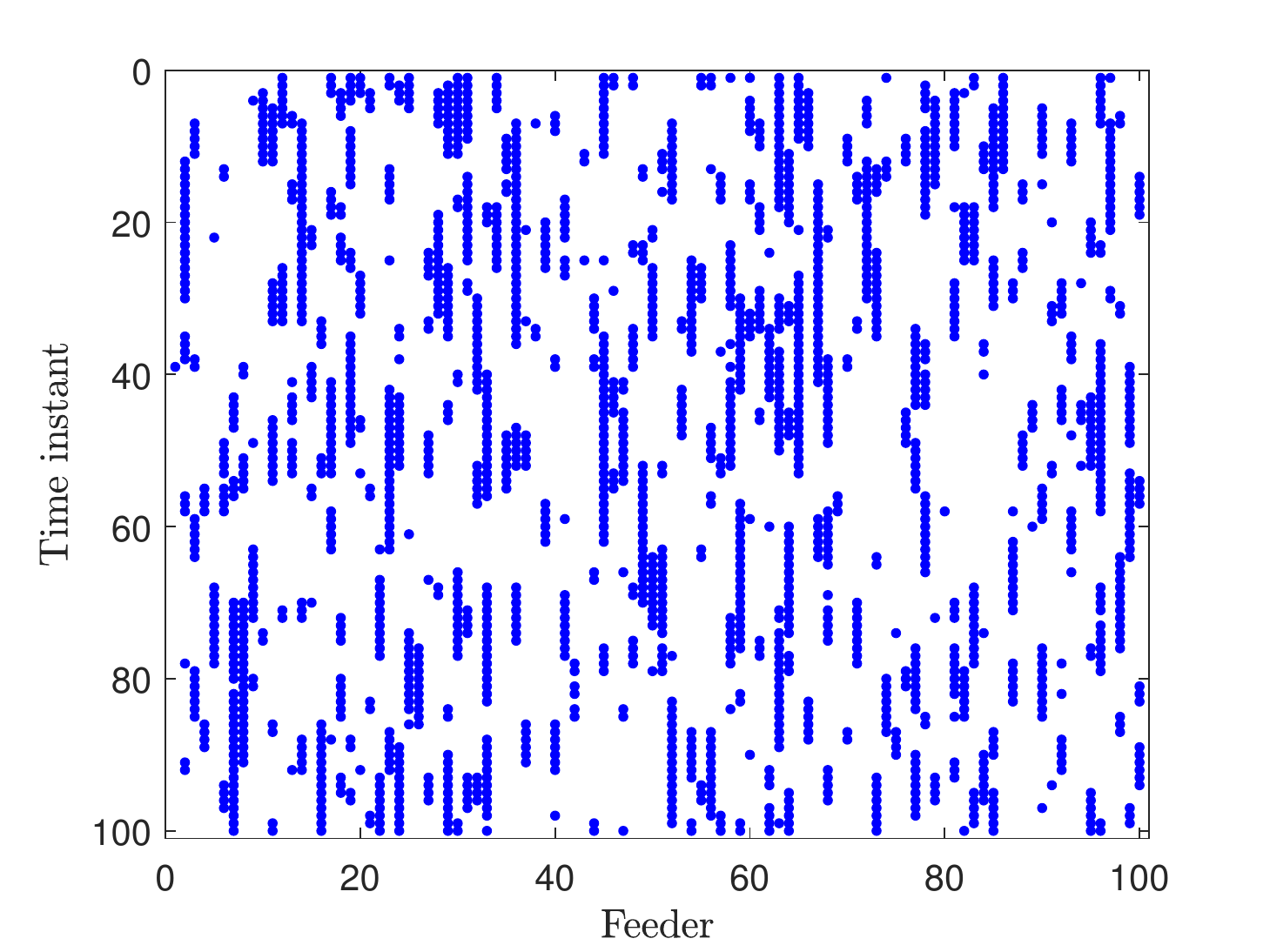}
\caption{Positions of the observed entries, $\Omega$, generated by the Markovian model for a $100 \times 100$ matrix, when $\mathbb{E}[L_0]=N$ and $\mathbb{E}[\gamma]=0.8$.}
\label{fig:mk_sampling}
\vspace{-5mm}
\end{figure}
\vspace{-1.5mm}
\subsection{Robustness with respect to different sampling patterns}
The problem of recovering missing data when the subset of missing entries is not uniformly sampled is addressed in this section. In practical scenarios, a sensor failure or a downtime in the communication line provides the operator with a number of consecutive unavailable measurements in the state variable vectors. Let $L_0$ be the number of consecutive missing entries. The expected value of $L_0$ varies depending on the reliability of the sensing infrastructure. In the uniform sampling model this scenario is not possible. In contrast, a more general sampling procedure is introduced.

\begin{figure}[t!]
\centering
\includegraphics[width=3.7in]{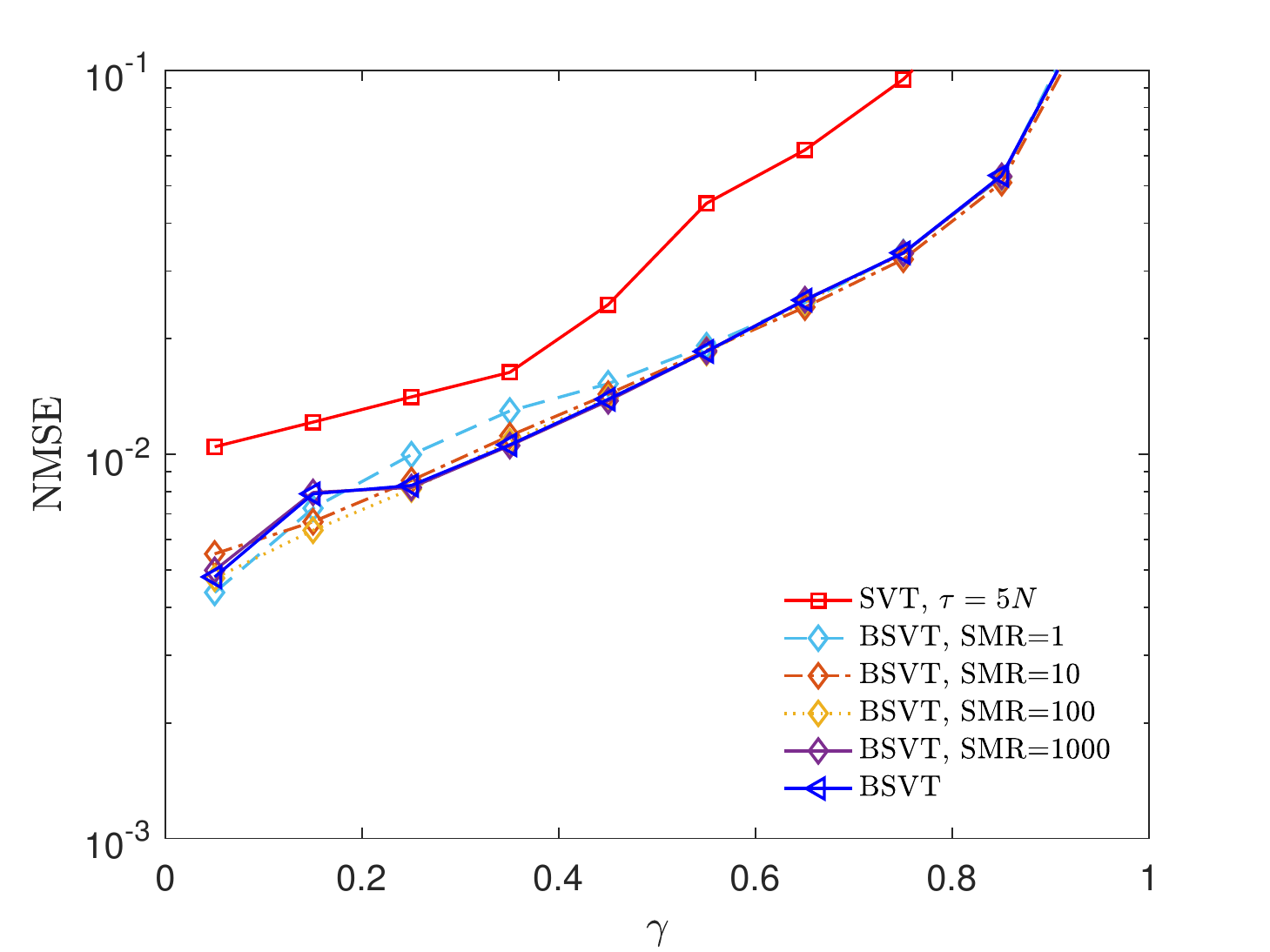}
\caption{Real data recovery performance for the Markov-chain-based sampling model, using SVT and BSVT for different levels of mismatch, when $\mathbb{E}[L_0]=N$ and $\textnormal{SNR}=20$ dB.}
\label{fig:mk_svt_bsvt}
\vspace{-3mm}
\end{figure}

\begin{figure}[t]
\centering
\includegraphics[width=3.3in]{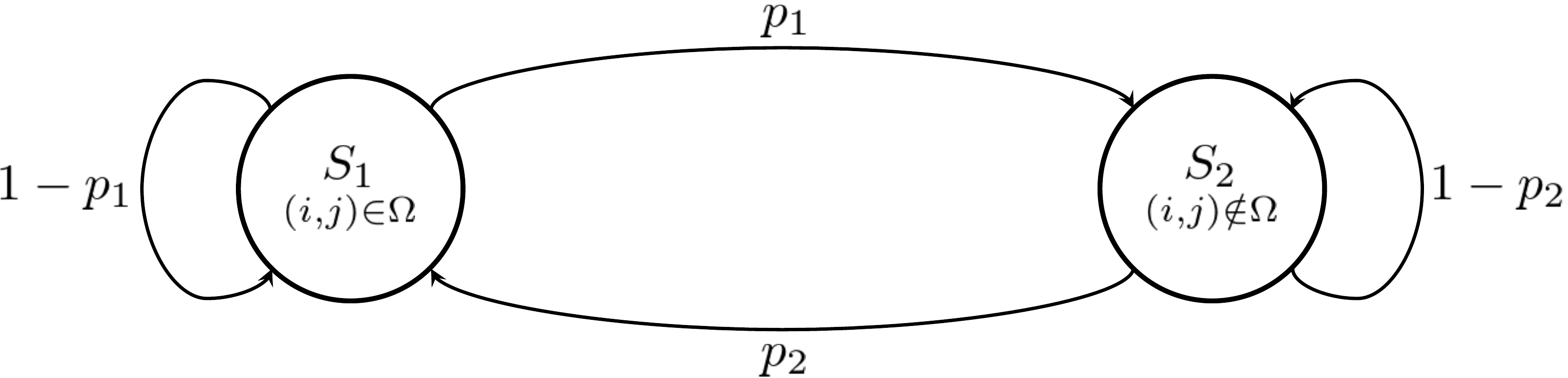}
\caption{State diagram for the Markovian sampling model.}
\label{fig:markov_model}
\vspace{-5mm}
\end{figure}

The proposed sampling model is based on a two-state Markov Chain. In this setting, for each entry $(\Mm)_{i,j}$ of the matrix $\Mm$, the finite state machine depicted in Fig. \ref{fig:markov_model}, is either in state $S_1$ in which case the entry $(i,j)$ is available to the operator, or in state $S_2$ in which case the entry is not available. As before, the set $\Omega$ contains all the entries from the matrix $\Mm$ that are available to the operator.
In Fig. \ref{fig:markov_model}, $p_1$ is the transition probability from state $S_1$ to $S_2$ and $p_2$ is the transition probability from $S_2$ to $S_1$. Hence, the expected value of the ratio of missing entries is given by the steady state probability of being in $S_2$. 
Consequently, the expected value of the ratio of missing entries for the Markovian sampling model is
\begin{equation} \label{eq:e_gamma}
\mathbb{E}[\gamma] = \frac{p_1}{p_1 + p_2}.
\end{equation}
The expected number of consecutive missing entries, $\mathbb{E}[L_0]$, is:
\begin{equation} \label{eq:el0_1}
\mathbb{E}[L_0] = \sum_{l=0}^{n} l \frac{p_1}{p_1 + p_2} (1-p_2)^l.
\end{equation}
Solving (\ref{eq:el0_1}) for $n \to \infty$ and combining with (\ref{eq:e_gamma}) leads to
\begin{equation} \label{eq:el0}
\mathbb{E}[L_0] =\frac{1-\mathbb{E}[\gamma]}{p_2^2} (1-p_2).
\end{equation}
Therefore, for any given $\gamma$ and $L_0$, using (\ref{eq:e_gamma}) and (\ref{eq:el0}), $p_1$ and $p_2$ are identified such that on average the sampling model in Fig. \ref{fig:markov_model} has a ratio of missing entries $\gamma$ and the length of the vectors with consecutive missing entries $L_0$.
Note that the case $\mathbb{E}[L_0]=1$ reduces to the uniform sampling model with probability $\mathbb{P}[(i,j) \in \Omega]=1-\gamma$.
In this framework, a comparison between the SVT and the BSVT-based recoveries is presented for the case in which the sampling pattern is not uniform. 
In order to consider the case in which a particular feeder does not provide any measurements, the expected length of the vectors with missing data is selected to be equal to the length of the state variable vectors, i.e., $\mathbb{E}[L_0]=N$.
Fig. \ref{fig:mk_sampling} shows an example of a sampling pattern generated by the Markov-chain-based model, when $\mathbb{E}[L_0]=N$ and $\mathbb{E}[\gamma]=0.8$.

Fig. \ref{fig:mk_svt_bsvt} compares the performance of the SVT-based recovery with the BSVT-based recovery for the case in which the matrix $\Mm$ is sampled using the Markov-chain-based sampling model with $\mathbb{E}[L_0]=N$. Different levels of mismatch are introduced to assess the robustness of the new algorithm to mismatched prior knowledge when the sampling pattern is not uniform. Remarkably, the performance of the proposed approach is not significantly affected by the amount of prior knowledge in any of the missing data regimes. Moreover, BSVT performs better than SVT when the sampling pattern is not uniform. A significant gain in performance is observed for small values of $\gamma$. Consider the following example for the sake of discussion, for a fixed tolerance of $10^{-2}$ in NMSE, the SVT algorithm recovers up to $4\%$ of the entries of the matrix $\Mm$ while BSVT recovers $40\%$ (See Fig. \ref{fig:mk_svt_bsvt}). The improvement in the data recovering performance for the same level of tolerance is significant.
Numerical results in this section show that BSVT is not only providing better performance than SVT when the entries are not uniformly sampled but it is also robust to mismatched statistics. The robustness of the new algorithm extends to different sampling patterns. In view of this, BSVT represents a better alternative for recovering missing data in practical scenarios than SVT and LMMSE estimation.

\section{Conclusion}
A novel algorithm for recovering missing data in data sets that admit a low rank description has been presented. The proposed approach, BSVT, combines the low computational cost of SVT with the optimality of the LMMSE estimator when the data source is modelled as a multivariate Gaussian random process and second order statistics are available. The combined new approach addresses the issues of individual recovery methods. The robustness of the new algorithm on both mismatched statistics and sampling patterns was demonstrated through simulations. In respect to the SVT algorithm the new approach addresses the issue of choosing the value of $\tau$ by calculating the optimal threshold at each iteration. Compared with the standard LMMSE estimator the new algorithm is robust to inaccurate second order statistics. In order to assess practical scenarios, a sampling model that incorporated missing state variable vectors, is illustrated.
The performance gain compared to SVT was significant for both uniform and non-uniform sampling models.
Ultimately, the proposed algorithm is shown to provide a robust and low complexity method to recover missing data in low voltage distribution systems.
\vspace{-1.5mm}
\bibliographystyle{IEEEbib}
\balance
\bibliography{references}

%
%
%

\end{document}